\documentclass[amssymb,aps,prb,letterpaper,twocolumn,showpacs]{revtex4-1}
\usepackage[utf8]{inputenc}
\usepackage{graphicx}
\usepackage{amsmath,amsfonts,amssymb}
\usepackage{physics}
\usepackage{multirow}
\usepackage{comment}
\usepackage{bbm}

\newcommand{\identity}{{\mathbbm{1}}}

\usepackage[dvipsnames]{xcolor}

\begin{document}
\title{Numerical evidence of a super-universality of the 2D and 3D random quantum Potts models}
\author{Valentin Anfray}
\author{Christophe Chatelain}
\affiliation{Universit\'e de Lorraine, CNRS, LPCT, F-54000 Nancy, France}
\date{\today}

\begin{abstract}
The random $q$-state quantum Potts model is studied on hypercubic lattices
in dimensions 2 and 3 using the numerical implementation of the Strong Disorder
Renormalization Group introduced by Kovacs and Igl\'oi
[Phys. Rev. B {\bf 82}, 054437 (2010)]. Critical exponents
$\nu$, $d_f$ and $\psi$ at the Infinite Disorder Fixed Point are estimated
by Finite-Size Scaling for several numbers of states $q$ between 2 and 50.
When scaling corrections are not taken into account, the estimates of both $d_f$
and $\psi$ systematically increase with $q$. It is shown however
that $q$-dependent scaling corrections are present and that the
exponents are compatible within error bars, or close to each other, when these
corrections are taking into account. This provides evidence of the existence of
a super-universality of all 2D and 3D random Potts models.
\end{abstract}
\maketitle

\section{Introduction}
Random quantum ferromagnets are known to undergo a very peculiar
phase transition for which quantum fluctuations, that drive the transition
in the absence of disorder, are dominated by disorder fluctuations. Thanks
to this peculiarity, the properties of the Infinite-Disorder Fixed Point (IDFP)
that governs the critical behavior at the transition can be studied using a
rather simple real-space renormalization group, introduced by Ma and
Dasgupta~\cite{MaDasgupta}, and referred to as Strong Disorder Renormalization
Group (SDRG). In the case of the random transverse-field Ising chain (RTIM),
Fisher was able to find the asymptotic solution of the flow
equations and determine exactly the critical exponents~\cite{Fisher1,Fisher2,
Monthus1,Monthus2}. In particular, the dynamical exponent $z$ is infinite at
the fixed point whereas $z=1$ for the pure RTIM. The excitation gap $\Delta E$
displays an essential singularity $\Delta E\sim e^{-aL^\psi}$ with the lattice
size $L$ where the critical exponent is $\psi=1/2$. The average magnetization
follows a power law $\mu\sim L^{d_f}$ with a fractal dimension $d_f=d-\phi\psi$
and a magnetic critical exponent $\phi$ equals to the golden
number $(1+\sqrt 5)/2$. These exponents are expected to be exact. Away from
the critical point, in the so-called Griffiths phase, the dynamics is dominated
by rare macroscopic clusters of strong (resp. weak) couplings that can order
earlier (resp. later) than the rest of the system~\cite{VojtaReview}. As a
consequence, the dynamical exponent is larger
than $1$ and diverges with the control parameter $\delta$ as $1/|\delta|$ as
the IDFP is approached~\cite{Igloi1}. Finally, the correlation length diverges
as $\xi\sim|\delta|^{-\nu}$ with $\nu=2$.
\\

The universality class of the RTIM turned out to be quiet robust: the number of states
$q$ of the 1D random quantum Potts model was shown to be an irrelevant parameter
in the SDRG flow equations~\cite{Senthil}. Therefore the critical behavior is described
by the same IDFP as the RTIM for all values of the number of states
$q$, in contrast to what is observed in the classical case where the magnetic
critical exponent increases smoothly with $q$~\cite{Jacobsen1,Chatelain1,Jacobsen2,Chatelain2}.
For a sufficiently strong disorder, the critical behavior of the random $q$-state
quantum clock model is also expected to be governed by the same IDFP as the
RTIM~\cite{Senthil,Carlon1}.
The random quantum $N$-color Ashkin-Teller chain, equivalent to $N$ coupled Ising
chains, has attracted much attention in the last decade. In the case $N=2$, the
phase diagram is qualitatively unchanged by the introduction of disorder~\cite{Carlon1}.
Along the self-dual transition line, the inter-chain coupling is an irrelevant parameter
in the SDRG flow equations~\cite{Goswani}. As a consequence, the critical behavior is again
the same as the random RTIM whereas exponents vary along the line in the pure case. For
strong inter-chain coupling, the transition line splits into two lines, enclosing a new
intermediate phase acting as a double-Griffiths phase~\cite{Vojta3,Chatelain3}. Despite
the fact that the inter-chain coupling flows towards an infinite value during
renormalization, the critical behavior is still in the RTIM universality class along
these two lines~\cite{Vojta3}. In the case $N\ge 3$, the pure Ashkin-Teller chain undergoes
a first-order phase transition, as the Potts model with $q\ge 4$, which becomes continuous
in presence of disorder~\cite{Goswani,Vojta1,Vojta2,Ibrahim,Chatelain4}. The critical
behavior is in the RTIM universality class at weak inter-chain coupling but seems to be
governed by a distinct IDFP at stronger inter-chain coupling~\cite{Vojta2}.
\\

In this paper, we address the question whether the robustness of the universality
class of the RTIM is specific to the 1D case or exists also in higher dimensions.
Much less is known about 2D or 3D random quantum ferromagnets. The extension of
the SDRG to higher dimensions is trivial but the flow equations are then too
complicated for an analytical solution to be found. The numerical implementation of
the SDRG rules is complicated by the fact that the topology of the lattice changes
during the renormalization. A single site, or cluster, is coupled to a large
number of other spins after only a few iterations. Nevertheless, the critical behavior
of the 2D RTIM could be shown to be governed by an IDFP~\cite{Motrunich}. An efficient
algorithm, allowing for accurate estimates of the critical exponents, was introduced
by Kovacs and Igl\'oi\cite{Kovacs1,Kovacs2}. They were able to show that the critical
exponents of the random RTIM depend on the dimension of the lattice.
Recently, the critical behavior of the 2D quantum Potts model with a quasi-periodic
modulation of the couplings was shown to be governed by an infinite quasi-periodicity
fixed point, distinct from an IDFP but with an infinite dynamical exponent like an
IDFP~\cite{Agarwal1,Agarwal2}. Interestingly, the critical exponents are compatible
within error bars for all numbers of states $q\ge 3$. Later on, Kang {\sl et al.}
discussed the IDFP of the random quantum Potts model in light of a mapping onto
a discrete gauge model where the size of the gauge group is equal to the number
of states $q$ of the Potts model~\cite{Vasseur}.
Quantum Monte Carlo simulations were performed for the two-dimensional random
quantum Ising model and the 3-state Potts model. The numerical estimates of the
critical exponents of the two models are in good agreement, providing evidence
of an independence on $q$ and the super-universality of the IDFP in two dimensions.
\\

In this work, the $q$-state random quantum Potts model is considered in two and three
dimensions. Since in the classical 2D random Potts model, the magnetic critical
exponent increases slowly with $q$, we considered number of states up to $q=50$.
Critical exponents are estimated numerically using the Kovacs-Igl\'oi
algorithm. In the first section of this paper, the model and the algorithm are
presented. The determination of the location of the critical points is detailed
in section III. The correlation length exponent $\nu$ is extracted from the
statistics of the pseudo-critical points. In section IV, the magnetic fractal
dimension $d_f$ is estimated from the Finite-Size Scaling of the average
magnetic moment. In section V, the exponent $\psi$ is estimated from the analysis
of the average energy gap. Conclusions follow.

\section{Potts model and SDRG algorithm}
\subsection{The random quantum Potts model}
The $q$-state quantum Potts model is defined on a lattice $\Lambda=(V,E)$ by the
Hamiltonian~\cite{Stefen,Solyom}
\begin{equation}
    H_{\rm Potts}=-\sum_{(i,j)\in E} J_{ij}D_{i,j}-\frac{1}{q}\sum_{i\in V} h_i M_i
    \label{HamiltonienP}
\end{equation}
acting on the Hilbert space spanned by the states $\bigotimes_{i\in V}\ket{n_i}$ with $n_i=0,\ldots,q-1$.
The first sum extends over the set $E$ of edges of the lattice $\Lambda$ and the matrix elements
of the diagonal operator $D_{i,j}$ vanish unless $n_i=n_j$, in which case they are equal to 1.
A representation of this operator in terms of local operators is given by
   \begin{equation}
       D_{i,j}={1\over q}\sum_{n=0}^{q-1}\Omega_i^n\Omega_j^{-n}
   \end{equation}
where $\Omega_i=\identity^{\otimes i-1}\otimes\Omega\otimes\identity^{\otimes N-i}$
($N$ is the number of sites of the lattice) and $\Omega$ is a diagonal $q\times q$ matrix whose
diagonal elements are $\omega^n$ with $\omega=e^{2i\pi/q}$. In the pure case, i.e. $J_{ij}=J>0$, the first
term of the Hamiltonian favors a ferromagnetic ordering of the spins, i.e. $n_i=n$ $\forall i$.
The second sum of the Hamiltonian~(\ref{HamiltonienP}) extends over the set $V$ of sites of the
lattice $\Lambda$ and $M_i=\identity^{\otimes i-1}\otimes M\otimes\identity^{\otimes N-i}$ with
$M$ the $q\times q$ matrix whose elements are all equal to 1. In the pure case, $h_{i}=h$, the
second term of the Hamiltonian destroys the ferromagnetic ordering and is associated to quantum
fluctuations. When $q=2$, the Potts model is equivalent to the RTIM whose Hamiltonian takes the
simpler form
\begin{equation}
    H_{\rm Ising} = - \sum_{(i,j)\in E} J_{ij} \sigma_i^z \sigma_j^z
    - \sum_{i\in V} h_i \sigma_i^x
\end{equation}
where $\sigma_i^{x,z}$ are Pauli matrices acting on the site $i$ of the lattice.
In the one-dimensional case, only the Ising model is exactly solvable when $J_{ij}=J$ and $h_i=h$.
Duality arguments predict that the transition point is located at $J=h$ for any number of states
$q$. The pure Potts chain undergoes a second-order phase transition when $q\le q_c(1)=4$ and a
first-order transition when $q>q_c(1)$. At higher dimensions $d>1$, the transition point is not
known exactly. The number of states $q_c(d)$ separating the regime of first and second-order phase
transition is also not known exactly for $d>1$.
\\

In the following, the Potts model with quenched disorder is considered. The
exchange couplings $J_{ij}$ and the transverse fields $h_i$ are independent
random variables distributed according to the distributions $P_0(J_{ij})$ and
$Q_0(h_i)$. As mentioned in the introduction, the critical exponents of the
random RTIM ($q=2$) has been determined exactly by Fisher.
In dimensions $d=2,3,$ and 4, they were estimated numerically by Kovacs and
Igl\'oi. In the following, the RTIM will be used as a test bed for our implementation of the
SDRG algorithm and for the analysis of the numerical data. In the regime $q>q_c(d)$, the
first-order phase transition of the pure Potts model is expected to be rounded by
disorder and turned into a continuous transition, as first discussed by Goswani {\sl et
al.}~\cite{Goswani}.
A rigorous proof was later given that an infinitesimal amount of disorder is sufficient
to round any first-order phase transitions in quantum systems in dimensions $d\le
2$~\cite{Aizenman1,Aizenman2}. For $d>2$, the first-order phase transition may survive
at weak disorder, as in the classical case for $d\ge 3$, for example in the random 3D
4-state classical Potts model~\cite{Chatelain5,Chatelain6}.
\\

In this work, several probability distributions were considered. The uniform
distribution
 \begin{equation}
    P_0(J_{ij})=\Theta(J_{ij})\Theta(1-J_{ij}),
    \label{ProbaP}
  \end{equation}
where $\Theta$ is the Heaviside function, for the exchange couplings and
  \begin{equation}
    Q_0(h_i) = \frac{1}{h_{\rm max}}\Theta(h_i)\Theta(h_{\rm max}-h_i)
    \label{ProbaQ}
  \end{equation}
for the transverse fields. The Potts model is in the ferromagnetic phase
for a sufficiently small parameter $\theta=\log h_{\rm max}$ and in the
paramagnetic phase for large $\theta$. The distributions (\ref{ProbaP})
and (\ref{ProbaQ}) are referred to as weak disorder in the
following. These distributions are expected to evolve along the RG flow
and become broader and broader. Because the distributions $P_0$ and
$Q_0$ are far from the distributions at the IDFP, corrections to scaling
are expected for small lattice sizes and therefore small numbers of RG steps.
To minimize these corrections, pow-law distributions
	\begin{eqnarray}
	&&P_0(J_{ij})\sim J_{ij}^{-\Delta},\quad (0<J_{ij}<1)\nonumber\\
	&&Q_0(h_i)\sim h_i^{-\Delta},\quad (0<h_i<h_{\rm max})
	\label{Proba2}
	\end{eqnarray}
were also considered for several numbers of states $q$ of the Potts model.
The value $\Delta=2/3$ is referred to as a medium disorder and $\Delta=4/5$
as a strong disorder.   

\subsection{SDRG algorithm}
The Strong Disorder Renormalization Group (SDRG) is a real-space decimation scheme
where the strongest coupling $\Omega = \max(\{J_{ij}\},\{h_i\})$ is decimated at each
iteration~\cite{Fisher1,Fisher2,Monthus1}. The case of the RTIM is discussed first. If the
strongest coupling is an exchange coupling, say $\Omega = J_{ij}$, the two spins $i$ and $j$ are
merged into a new effective cluster whose magnetic moment $\mu' = \mu_i + \mu_j$ is
the sum of the moments $\mu_i$ and $\mu_j$ of the two spins $i$ and $j$. Second-order perturbation
theory shows that this new cluster is coupled to an effective transverse field
$h'=\frac{h_ih_j}{J_{ij}}$ and to any other spin $k\ne i,j$ by an exchange coupling
$J_{ik}'=J_{ik}+J_{jk}$. If the strongest coupling $\Omega$ is a transverse field,
say $h_i$, the spin $i$ is decimated. An effective coupling is induced between all pairs
of spins $k$ and $l$ that were both coupled to site $i$. To second order in perturbation theory,
this effective coupling is $J'_{kl} = J_{kl} + \frac{J_{ik}J_{il}}{h_i}$.  As this scheme is
iterated, the probability distributions $P(J)$ and $Q(h)$ of the couplings become broader and
broader so that second-order perturbation theory is expected to become exact
at the IDFP. The sum rule can then
be replaced by a maximum rule: at the IDFP, it is sufficient to write the exchange coupling of a
spin $k$ with the new effective cluster at site $i$ as the maximum $J_{ik}'=\max(J_{ik},
J_{jk})$ instead of the sum. Similarly, the effective exchange coupling $J_{kl}'$
induced by the decimation of the site $i$ can be simplified as $J'_{kl} = \max(J_{kl},
\frac{J_{ik}J_{il}}{h_i})$. For the Potts model, the SDRG rules are shown to be~\cite{Senthil}
    \begin{equation}
        h' = \frac{h_ih_j}{\kappa J_{ij}},\quad
        J'_{kl} = J_{kl} + \frac{J_{ik}J_{il}}{\kappa h_i}
        \label{SDRGrules}
    \end{equation}
where $\kappa = q/2$. These rules are the same for any dimension $d$ of the lattice. However,
the main difficulty in implementing them numerically when $d>1$ comes from the increasing
number of couplings $J_{ij}$ that are generated at each decimation. Finding the largest coupling
requires more and more CPU time and even storing all the couplings restricts the application to
small lattice sizes. A crucial simplification was introduced by Kovacs and
Igl\'oi~\cite{Kovacs1,Kovacs2}. They showed that many couplings are actually irrelevant
at the IDFP. The resulting algorithm and the details of our implementation
for the Potts model are discussed in the following.
\\

The Hamiltonian can be seen as a weighted graph. A weight
\begin{equation}
    r_i = -\ln h_i
    \label{Algo1}
\end{equation}
is attached to each node $i$ and an edge with a distance
\begin{equation}
    d_{ij} = d_{ji} =- \ln J_{ij}
    \label{Algo2}
\end{equation}
is defined between each pair $(i,j)$ of nodes of the graph for which $J_{ij}\ne 0$.
With the above definitions, the SDRG rules become:
\begin{itemize}
    \item
    If $r_i$ is the global minimum, equivalently if $h_i$ is the global maximum, then the
    node $i$ is removed. For all pairs of sites $(k,l)$ connected to $i$, i.e. such that
    $J_{ik},J_{il}\ne 0$, the distance $d_{kl}$ is updated as
       \begin{equation}
           d_{kl}'=\min(d_{kl},d_{ki}+d_{il}-r_i+\ln\kappa)
       \end{equation}
    which is equivalent to (\ref{SDRGrules}) when replacing the sum rule by the maximum rule.
    \item
    If $d_{ij}$ is the global minimum, or equivalently if $J_{ij}$ is the global maximum, then the
    two nodes $i$ and $j$ are merged into a single node $i$ whose weight is
    \begin{equation}
        r_i' = r_i + r_j - d_{ij} + \ln \kappa
    \end{equation}
    which is again equivalent to (\ref{SDRGrules}). For each node $k$ previously
    connected to both $i$ and $j$, i.e. $J_{ik},J_{jk}\ne 0$, the distance to
    the new site $i$ is updated as $d_{ki}'=\min(d_{ik}, d_{jk})$.
\end{itemize}

Several improvements can be implemented. First, instead of removing the node $i$ when the spin is
decimated ($\Omega=h_i$), it is set as inactive. The definition (\ref{Algo2}) of the distances
$d_{ij}$ are modified to
\begin{equation}
    d_{ij} = d_{ji} =- \ln J_{ij} + \frac{l_i}{2}\ln(\kappa h_i) + \frac{l_j}{2}\ln(\kappa h_j)
    \label{Algo2b}
\end{equation}
where $l_i$ is the activation status of the node $i$ which takes the value $l_i = 0$ when the node
has not been decimated yet (the node is then said to be active) and 1 otherwise (the node is
inactive). By setting the node $i$ as inactive instead of removing it when $\Omega=h_i$, it is
not necessary anymore to add new edges associated to the effective couplings that are generated
by second order perturbation theory. Instead, the exchange coupling between two active sites $k$
and $l$ is computed {\sl on the fly} when necessary as $J_{kl}=e^{-\delta_{kl}}$ where
$\delta_{kl}$ is the shortest distance of all paths of the graph joining sites $k$ and $l$ and
going through inactive sites only. If no path connects $k$ and $l$ then $J_{kl}=0$. The condition
of the shortest distance is equivalent to the maximum rule and the SDRG rule is recovered. Indeed,
when the site $i$ is decimated, $\delta_{kl}=d_{ki}+d_{il}=- \ln J_{ki} + \ln(\kappa h_i) - \ln
J_{il}$ so that the effective exchange coupling is $J_{kl}'=e^{-\delta_{kl}}={J_{ki}J_{il}\over
\kappa h_i}$ as expected. The advantage of this implementation is that the number of edges does not grow
during site decimation. Inactive sites are removed during edge decimation: if $\Omega
=\delta_{ij}$, sites $i$ and $j$ are merged into a new cluster on site, say $i$. All
inactive sites $k$ belonging to the shortest path between sites $i$ and $j$ can now
be removed and all edges $d_{kl}$ are added to $d_{il}$ using the minimum rule
$d_{il}'=\min(\delta_{il},d_{ik}+d_{kl})$.
\\

The computation of the shortest distance $\delta_{kl}$ between two sites can be time-consuming.
Hopefully, the shortest distance can be determined efficiently using Dijkstra
algorithm~\cite{Djikstra}. Note that Dijkstra algorithm requires the distances to be positive.
The distance $d_{ij}=-\ln J_{ij}$ between two active sites can be made positive by initially
choosing all exchange couplings $J_{ij}$ smaller or equal to 1. When $\kappa\ge 1$, the SDRG
rules imply that $J_{ij}'\le 1$ after renormalization. If the site $j$ is inactive while $i$ is
active, the distance $d_{ij}=-\ln J_{ij}+{1\over 2}\ln(\kappa h_j)$ is positive too because
the decimation of the node $j$ has been possible only if $h_j>J_{ij}$. It follows that $-\ln
J_{ij}+{1\over 2}\ln h_j+{1\over 2}\ln\kappa>-{1\over 2}\ln J_{ij}+{1\over 2}\ln\kappa\ge 0$
which completes the proof that $d_{ij}\ge 0$ when $\kappa\ge 1$ and $J_{ij}\le 1$.
However, the distance $d_{ij}$ can be negative when the two sites are
inactive. When it is the case, any path reaching site $i$ will then go to site $j$.
As a consequence, for any neighbor $k\ne i$ of site $j$, one can create or update
the distance between $i$ and $k$ as $d_{ik}'=\min(d_{ik},d_{ij}+d_{jk})$ and remove
the site $j$ and all edges $d_{jk}$.
\\

The second improvement concerns the choice of the next coupling to be renormalized.
It is not necessary to find the global minimum. Finding and decimating a local minimum is
sufficient if the local minimum is defined as:
\begin{itemize}
    \item $r_i$ is a local minimum if $r_i < \delta_{ik}$ for all $k$ such that there exists
    at least one path between $i$ and $k$. No edge involving the node $i$ can therefore be
    decimated before the node $i$.
    \item $\delta_{ij}$ is a local minimum if $\delta_{ij} < r_i, r_j$ and if $\delta_{ij} <
    \delta_{ik}$ or $\delta_{ij}<\delta_{jk}$ when there exists at least one path between $i$ or $j$
    and $k$.
\end{itemize}
These definitions ensure that a local minimum remains a local minimum when any another node or edge
is decimated first. The proof follows from the fact that:
\begin{itemize}
    \item
    if the node $j$ is decimated first, which implies that $r_j<\delta_{jk}$ for all sites $k$
    for which there exists at least one path joining $j$ and $k$, the site $j$ is set inactive.
    The distances between the site $j$ and its neighbors $l$ are updated to $d_{jl}'=d_{jl}
    -{1\over 2}r_j+{1\over 2}\ln\kappa$. The shortest distance $\delta_{ik}$ is therefore
    unchanged if the shortest path does not go through the site $j$ and becomes
    $\delta_{ik}'=\delta_{ij}+\delta_{jk}-r_j+\ln\kappa$ otherwise. Since $r_j<\delta_{jk}$
    and $r_i<\delta_{ij}$ if $r_i$ is a local minimum, the new value $\delta_{ik}'$
    is necessarily larger than $r_i$. $r_i$ remains therefore a local minimum.

	\item
    if the edge $\delta_{jk}$ is decimated first, the distance between a site $i\ne j,k$
    and the new site, say $j$, is updated to the value $\delta_{ij}'=\min(\delta_{ij},\delta_{ik})$.
    If $r_i$ is a local minimum, the condition $r_i<\delta_{ij},\delta_{ik}$ holds. It follows
    that $r_i<\delta_{ij}'$ and therefore, $r_i$ remains a local minimum.
\end{itemize}
Similarly, the distance $\delta_{ij}$ remains a local minimum in the following situations:
\begin{itemize}
    \item
    if the node $k$ is decimated first, which implies that $r_k<\delta_{kl}$, for all sites $l$
    for which there exists at least one path joining $k$ and $l$, the shortest distance
    $\delta_{ij}$ is unchanged if the shortest path does not go through the site $k$ and
    becomes $\delta_{ij}'=\delta_{ik}+\delta_{jk}-r_k+\ln\kappa$ otherwise. Since
    $r_k<\delta_{ik},\delta_{jk}$ and $\delta_{ij}<\delta_{ik}$ or $\delta_{ij}<\delta_{jk}$
    if $\delta_{ij}$ is a local minimum, the effective distance $\delta_{ij}'=\delta_{ik}
    +\delta_{jk}-r_k+\ln\kappa$ is larger than $\delta_{ij}$. The latter will therefore
    remain a local minimum.

    \item
    if the edge $\delta_{kl}$ is decimated first, the distance between a site $i\ne k,l$
    and the new site, say $k$, is updated to the value
    $\delta_{ik}'=\min(\delta_{ik},\delta_{il})$.
    If the $\delta_{ij}$ was a local minimum with the condition
    $\delta_{ij}<\delta_{ik},\delta_{il}$
    then $\delta_{ij}<\delta_{ik}'$ and therefore $\delta_{ij}$ remains a local minimum.

    \item
    if the edge $\delta_{jk}$ is decimated first, the weight of the new node $j$ becomes
    $r'_j=r_j+r_k-\delta_{jk}+\ln\kappa$. Since $\delta_{jk}<r_k$, the inequality $r_j'>r_j$
    holds. Therefore, if $\delta_{ij}$ is a local minimum, the condition $\delta_{ij}<r_j$ is
    preserved when the edge $\delta_{jk}$ is decimated. Distances will also be modified.
    In particular, $\delta_{ij}$ will be replaced by $\min(\delta_{ij},\delta_{ik})$.
    Since $\delta_{jk}$ was decimated first, $\delta_{jk}<\delta_{ij}$ so the condition
    $\delta_{ij}<\delta_{il}$ for all $l$ for which there exists a path between $i$ and
    $l$, should hold for $\delta_{ij}$ to be local minimum.
    In particular, $\delta_{ij}<\delta_{ik}$ and therefore the value of $\delta_{ij}$
    will not be modified by the decimation of $\delta_{jk}$. For $l\ne k$, $\delta_{il}$
    is unchanged because a path joining $i$ and $l$ can only go through inactive sites,
    so neither $j$ or $k$. In conclusion, $\delta_{ij}$ will remain a local minimum after
    the decimation of $\delta_{jk}$.
\end{itemize}

Local minimum can be decimated in any order. One can check that the same
decimations as in the original SDRG will take place. Only the order differs.
A lot of computation time is saved in looking for local minimum instead of
the global one. The drawback of this method is that the renormalization
flow being modified, one cannot study anymore the evolution of the total
magnetic moment or the number of sites as a function of $\Omega$.
Instead, the critical exponents should be estimated from the behavior of
the magnetic moment or the transverse field of the last decimated site
in the original SDRG. In our implementation, this site could have been
decimated anywhere in the RG flow so one has to keep track of the smallest
transverse field at each decimation.

\section{Critical point}\label{sec3}
\begin{figure}
    \centering
    \includegraphics[width=0.47\textwidth]{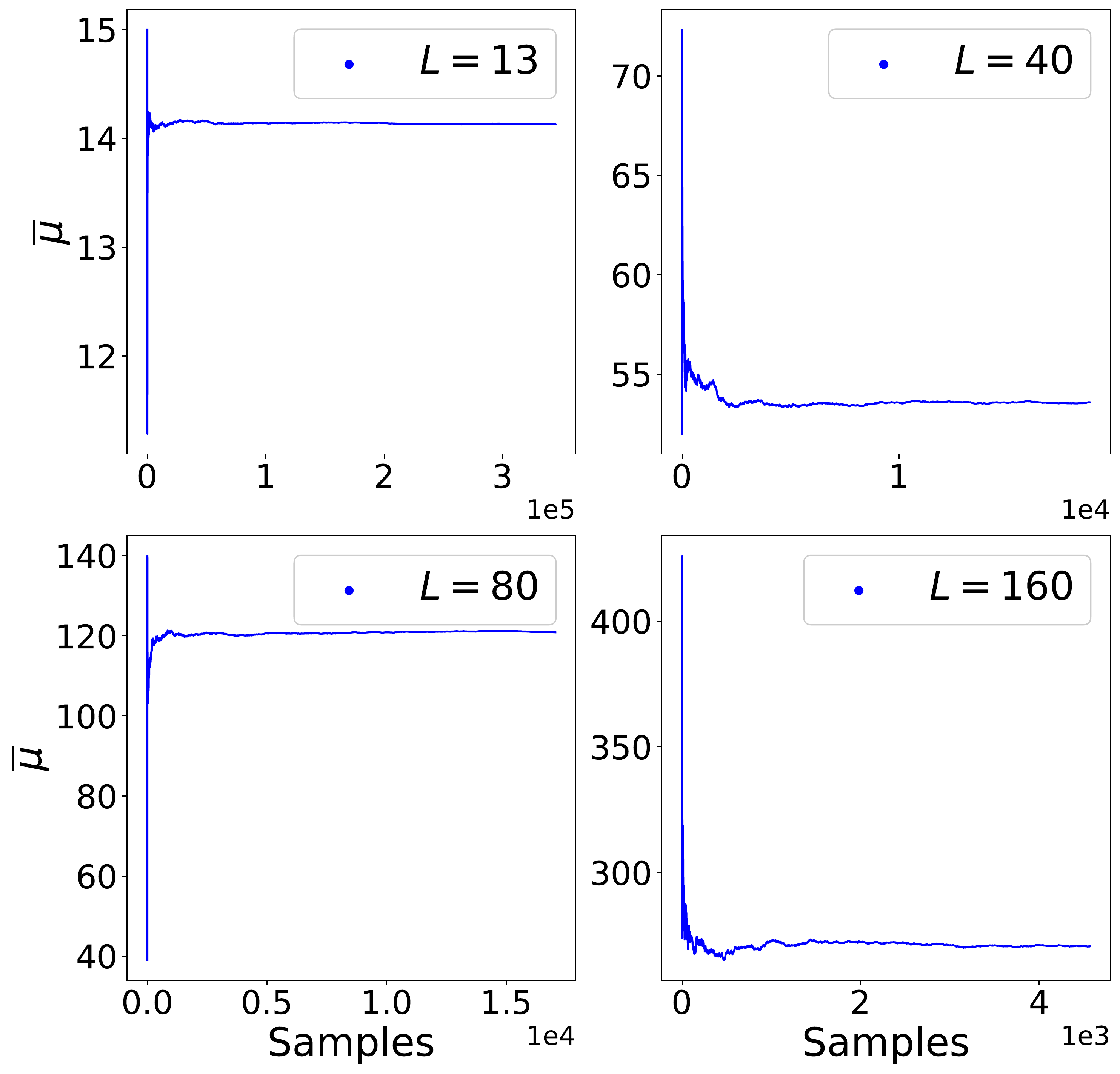}
    \caption{Average value of the magnetic moment $\mu$ of the last decimated cluster
    of the 3D $q=10$-Potts model with respect to the number of disordered samples.
    The four graphs correspond to different lattice sizes $L$. Note that the number
    of samples given on the $x$-axis has to be multiplied by $10^5$ ($L=13$, top left),
    $10^4$ ($L=40$, top right), $10^4$ ($L=80$, bottom left) and $10^3$ ($L=160$, bottom right).}
    \label{fig1}
\end{figure}

The random $q$-state Potts model is studied on 2D and 3D hypercubic lattices with
the above-detailed algorithm. The data have been averaged over more than $3000$
disordered configurations for the largest lattice sizes and up to $10^6$ for the
smallest ones. These numbers were chosen in order to achieve a good convergence
of average quantities. On Fig.~\ref{fig1}, the average magnetic moment of the last
decimated cluster during the original SDRG is plotted versus
the number of samples in the case of the 3D 10-state Potts model. Rare events
with a large contribution, usually expected in random systems, do not seem to
have any influence on the plateau reached by the average magnetic moment.
As can be seen on Fig.~\ref{fig1} in the case of the 3D 10-state Potts model,
the relative fluctuations of $\bar\mu$ are, in the worst case $L=160$,
of order ${\Delta\mu\over\bar\mu}\simeq 1/150<1\%$. The estimation of the error
as $\sqrt{{\rm Var}\ \!\mu/N}$, where ${\rm Var}\ \!\mu$ is the variance of the
data and $N$ the number of disordered configurations, leads to a relative error
of $0.6\%$ for the 3D 10-state Potts model at $L=160$. For the critical exponents
that will be estimated in the following, the error due to the finite
number of disordered configurations is a small contribution compared to the
error coming from the fits.

A pseudo-critical point $\theta_c^i(L)\equiv \ln h_c^i(L)$ is determined for
each disordered sample $i$ using the doubling method~\cite{Kovacs1,Kovacs2}.
Two identical replicas of the same system, i.e. with the same exchange couplings
and transverse fields, are glued together with some specific boundary conditions.
When the SDRG procedure is applied to both the joint system of size $2L$ and the
initial one of size $L$, the ratio $\mu(2L)/\mu(L)$ of the magnetic moments of
the last decimated cluster is expected to show a jump at the pseudo-critical point.
In the paramagnetic phase, $\theta>\theta_c^i(L)$, the decimated cluster is
located in one of the two replicas and $\mu(2L)=\mu(L)$. In contrast, in the
ferromagnetic phase, $\theta<\theta_c^i(L)$, the last decimated cluster spans
the two replicas and $\mu(2L)=2\mu(L)$. In practise, the system is considered
to be in the ferromagnetic phase if the last decimated cluster contains the same
sites in both replicas. To locate the pseudo-critical point, an interval
$[\theta_1,\theta_2]$ is manually chosen and is refined by performing additional
simulations at $\theta={1\over 2}(\theta_1+\theta_2)$ until the targeted
accuracy $\epsilon=|\theta_2-\theta_1|$ is reached. In the following,
the accuracy on the pseudo-critical point $\theta_c^i(L)$ is $10^{-5}$.
\\

\begin{figure}
    \centering
    \includegraphics[width=0.47\textwidth]{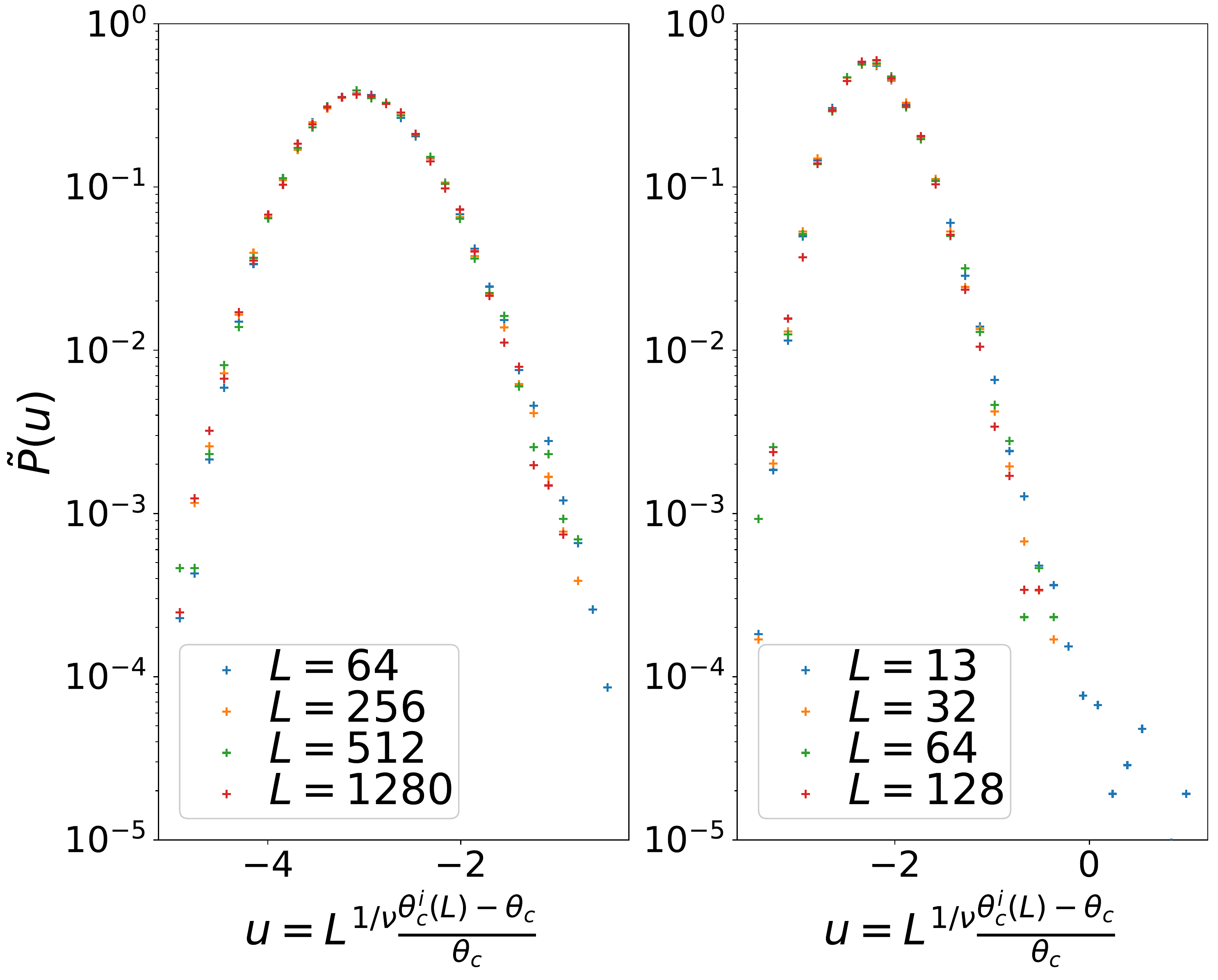}
    \caption{Probability distribution $P(u)$ of the rescaled distance
    $u=L^{1/\nu}(\theta_c^i(L)-\theta_c)/\theta_c$ to the critical point
    for the four largest lattice sizes that were considered in the
    minimization of the cost function Eq.~\ref{eq2}.
    On the left, the data for the $2D$ 10-state Potts model are plotted
    with the parameters $\nu = 1.2594$ and $\theta_c = 1.457$. On the right,
    the data for the $3D$ 10-state Potts model are plotted with the parameters
    $\nu = 1.008$, $\theta_c=2.332$.}
    \label{fig2}
\end{figure}

As the lattice size is increased, the pseudo-critical points $\theta_c^i(L)$
are expected to converge to the critical point $\theta_c$ of the infinite system as
   \begin{equation}
       |\theta_c^i(L)-\theta_c|\sim L^{-1/\nu}
       \label{eq1}
   \end{equation}
where $\nu$ is the correlation length exponent. As a consequence, the probability distribution
of the pseudo-critical points $\theta_c^i(L)$ should be independent of the lattice size $L$
when plotted with respect to the rescaled distance to the critical point
$u=L^{1/\nu}|\theta_c^i(L)-\theta_c|/\theta_c$. This plot is shown on Fig.~\ref{fig2} for
the 2D and 3D 10-state Potts models. As expected, all points fall nicely on the same curve
when the two parameters $\theta_c$ and $\nu$ are appropriately chosen. To determine the values
leading to the best collapse of the probability distributions $P_L(u)$ for different lattice
sizes $L$, the following cost function
\begin{equation}
    \sigma = \frac{1}{u_{\rm max} - u_{\rm min}}\int_{u_{\rm min}}^{u_{\rm max}}
    \sum_L \Big[P_L(u)-{1\over N_L}\sum_L P_L(u)\Big]^2du
    \label{eq2}
\end{equation}
was numerically minimized using Powell method. Only $N_L=4$
lattice sizes were considered. $u_{\rm min}$ and $u_{\rm max}$ are the smallest
and largest values of $u$ in the dataset. The optimal values of the parameters
$\theta_c$ and $\nu$ are collected in Table~\ref{tab1} for the 2D and 3D Potts
models. The estimates of the correlation length exponent $\nu$ are compatible
within error bars for the 2D (resp. 3D) Potts model, independently of
the number of states $q$ of the Potts model.

\begin{table}
\caption{\label{tab1}
Estimates of the critical point $\theta_c$ and correlation length exponent
$\nu$ of the 2D and 3D random Potts models. $\theta_c$ and $\nu$ were obtained
by imposing the collapse of the probability distribution $P(u)$.
}
\begin{ruledtabular}
\begin{tabular}{l|llllll}
2D & $q=2$ & $3$ & $5$ & $10$ & $20$ & $50$\\
\hline
$\theta_c$ & 1.678(1) & 1.563(1) & 1.495(1) & 1.457(1) & 1.442(1) & 1.435(1)  \\
$\nu$ & 1.25(2) & 1.24(2) & 1.25(2) & 1.26(2) & 1.24(2) & 1.25(2) \\
\hline
3D & $q=2$ & $3$ & $5$ & $10$ & $20$ & $50$\\
\hline
$\theta_c$ & 2.532(1) & 2.420(1) & 2.359(1) & 2.332(1) & 2.325(1) & 2.327(1) \\
$\nu$ &  1.01(1) & 1.004(10) & 1.001(10) &  1.008(10) & 1.004(10) & 1.005(10) \\
\end{tabular}
\end{ruledtabular}
\end{table}

\begin{figure}
    \centering
    \includegraphics[width=0.47\textwidth]{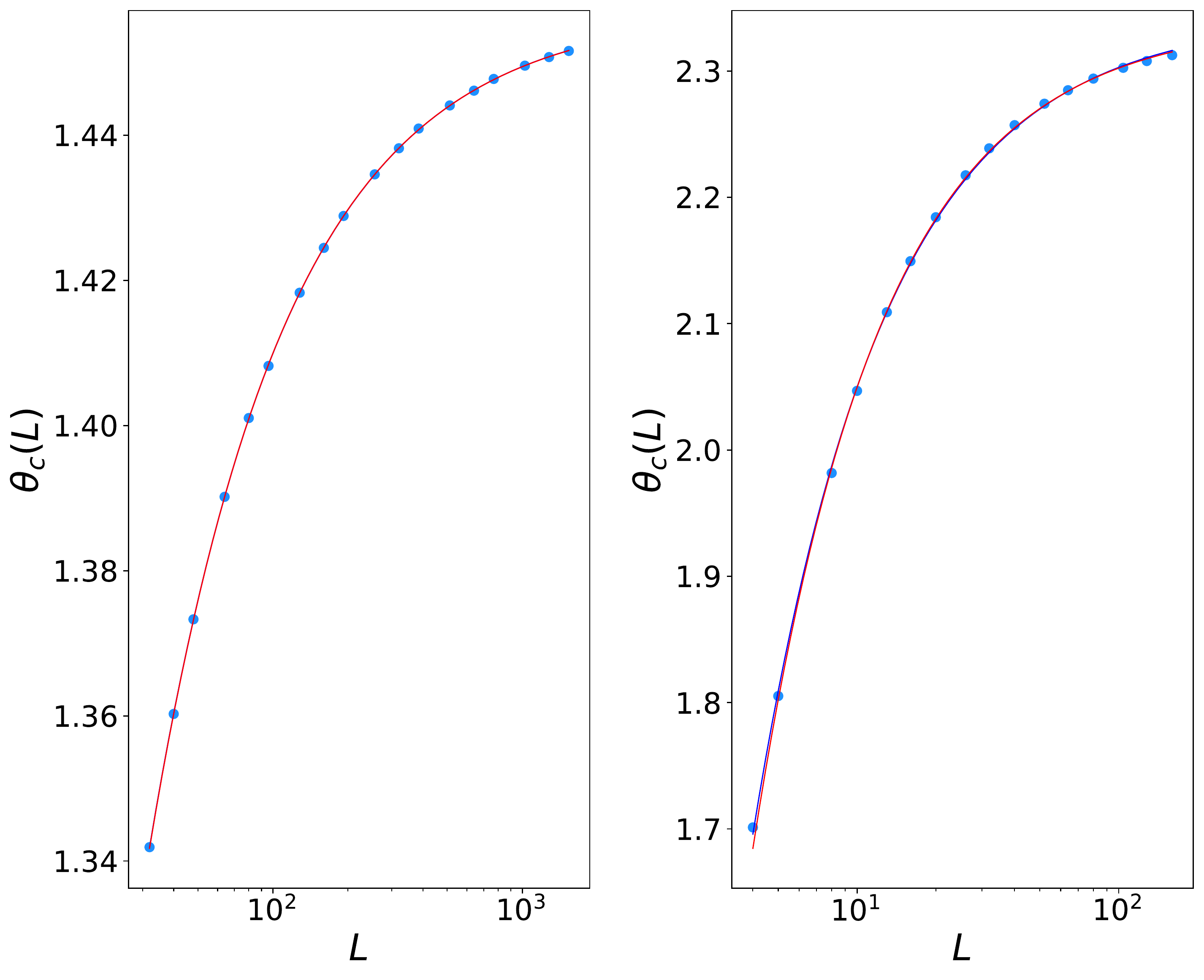}
	\caption{Average critical point $\bar\theta_c(L)$ versus the lattice
    size $L$ of the 2D (left) and 3D (right) $q=10$ Potts model.
	The blue curve corresponds to a simple fit without any correction
	while the red curve corresponds to a fit with an algebraic correction.
	They can be distinguished on the figure for small $L$ in the 3D case.}
	\label{fignus}
\end{figure}

\begin{figure}
    \centering
    \includegraphics[width=0.5\textwidth]{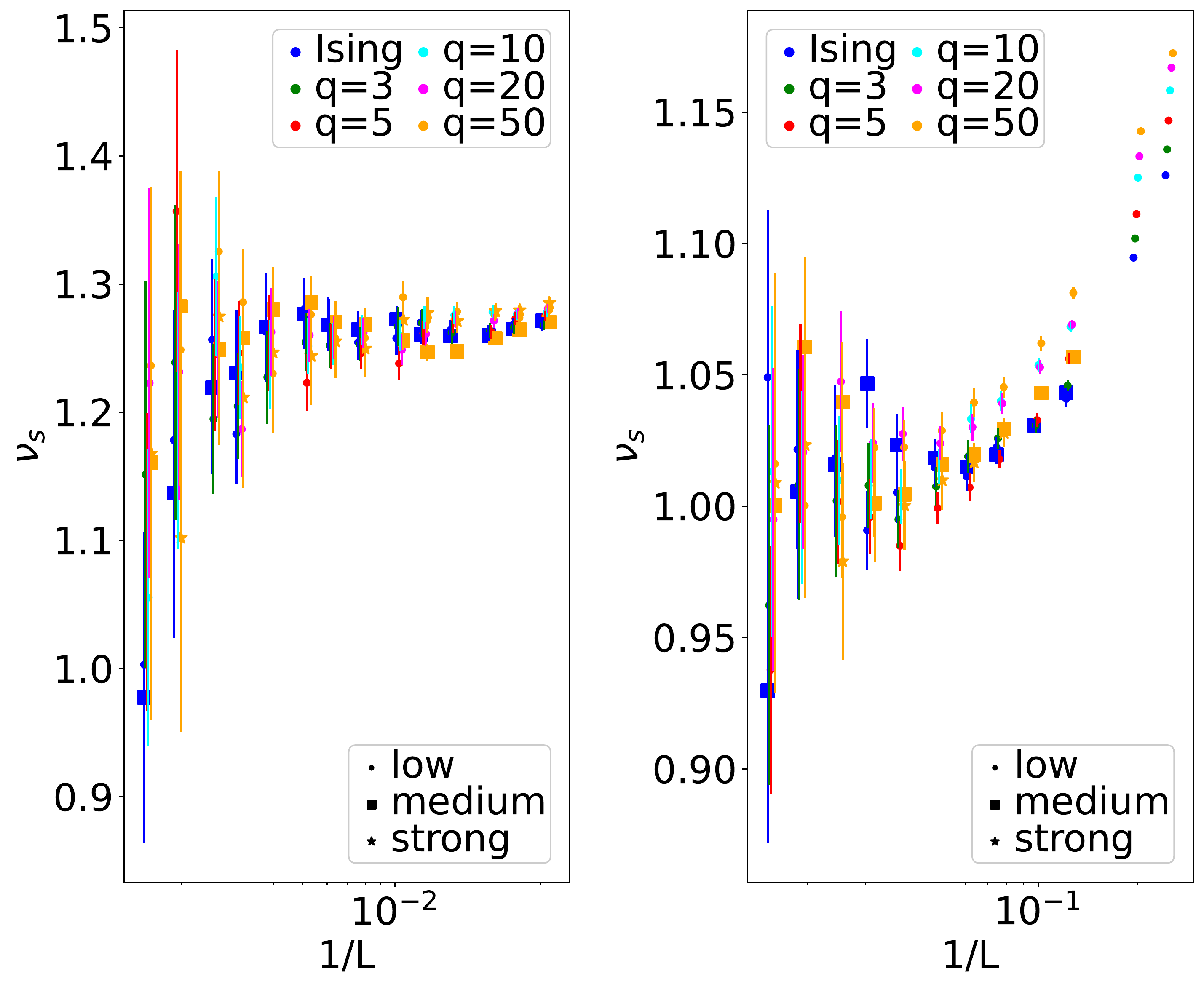}
    \caption{Effective exponent $\nu_s$ versus the smallest lattice size
    $L_{\rm min}$ entering into the fit for the 2D (left) and 3D (right)
    Potts models. The different colors correspond to different numbers of states
    $q$ of the Potts model. The different symbols correspond to different
    initial probability distributions of the couplings.}
    \label{fignus2}
\end{figure}

The correlation length exponent $\nu$ was also estimated by performing
a non-linear fit of the shift of the average pseudo-critical point
$\bar\theta_c(L)$ as $|\bar\theta_c(L)-\theta_c|=aL^{-1/\nu_s}$ where
$\theta_c$, $a$ and $\nu_s$ are free parameters. The notation $\nu_s$ is
used here to distinguish this estimate from the one obtained from the
collapse of the probability distribution. The average pseudo-critical
point is plotted on Fig.~\ref{fignus} for the 2D and 3D 10-state
Potts models. For the other values of the number of states $q$ that
were considered ($q=3, 5, 20$ and 50), the curves are similar.
We tried to take into account possible algebraic scaling corrections by
performing a non-linear fit of the data with the function $|\bar\theta_c(L)-\theta_c|
=aL^{-1/\nu_s}\big(1+bL^{-\omega}\big)$ where now $\theta_c$, $a$, $\nu_s$,
$b$ and $\omega$ are free parameters. Due to the small number of
degrees of freedom, this fit turned out to be quite unstable, different
fitting algorithms leading to incompatible values. An indirect method
was therefore applied to estimate the true exponent $\nu_s$: the non-linear
fit $|\bar\theta_c(L)-\theta_c|=aL^{-1/\nu_s}$ was performed on
various ranges of lattice sizes. A $L_{\rm min}$-dependent effective
exponent $\nu_s(L_{\rm min})$ is then estimated by a non-linear fit
restricted to the lattice sizes $L\ge L_{\rm min}$.
As can be seen in Fig.~\ref{fignus2} for $q=10$, this effective exponent
is relatively stable for small $L_{\rm min}$ but displays large fluctuations
at large $L_{\rm min}$ because of a number of degrees of freedom of the fit
becoming smaller and smaller. The error bars on $\nu_s(L_{\rm min})$ correspond
to the standard deviation of the fit. They do not take into account the
accuracy on $\theta_c(L)$, equal to $10^{-5}$, which leads to a much smaller
contribution (or order ${\cal O} (10^{-5})$) to the error on $\nu_s(L_{\rm min})$.
In the 2D case, the effective exponents do not vary significantly with
$L_{\rm min}$, which means that scaling corrections are weak.
The exponents for different numbers of states $q$ of the Potts model are
compatible within error bars. In contrast, for the 3D Potts model, the
influence of a correction is clearly seen as a stronger dependence of
the effective exponents on $1/L_{\rm min}$. For small $L_{\rm min}$,
the exponents $\nu_s$ of the $q$-state Potts models increase with $q$
and are fully incompatible. As $L_{\rm min}$ is increased, the exponents
take smaller values and their dispersion shrinks, although their fluctuations
increase. For large $L_{\rm min}$, a plateau seems to be reached around
$\nu_s\simeq 1.00$.

\begin{figure}
    \centering
    \includegraphics[width=0.47\textwidth]{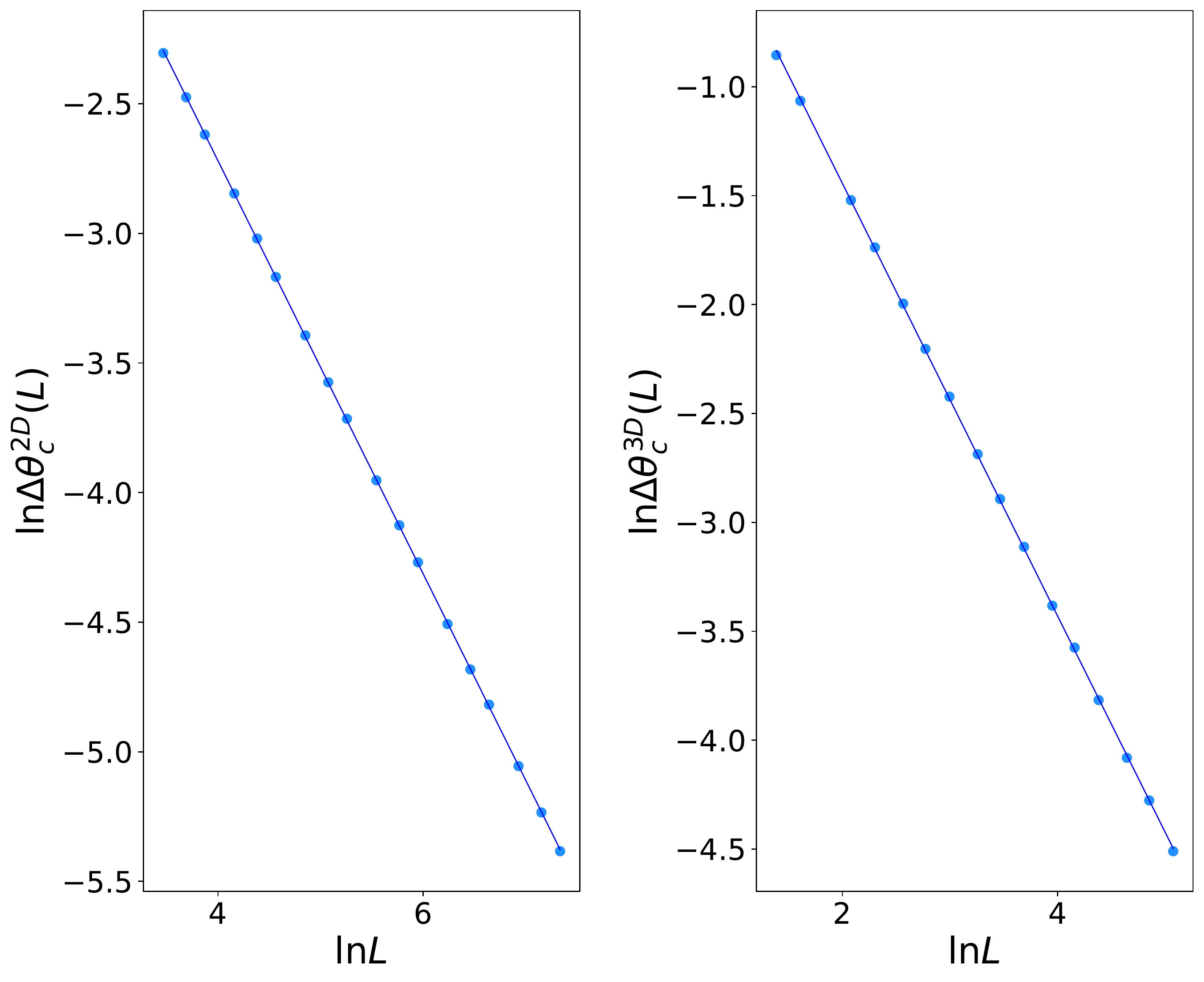}
    \caption{Standard deviation of the critical point $\Delta\theta_c(L)$ versus
    the lattice size $L$ of the 2D (left) and 3D (right) $q=10$ Potts models.
    The blue curve is a linear fit.}
    \label{fignuw}
\end{figure}

\begin{figure}
    \centering
    \includegraphics[width=0.5\textwidth]{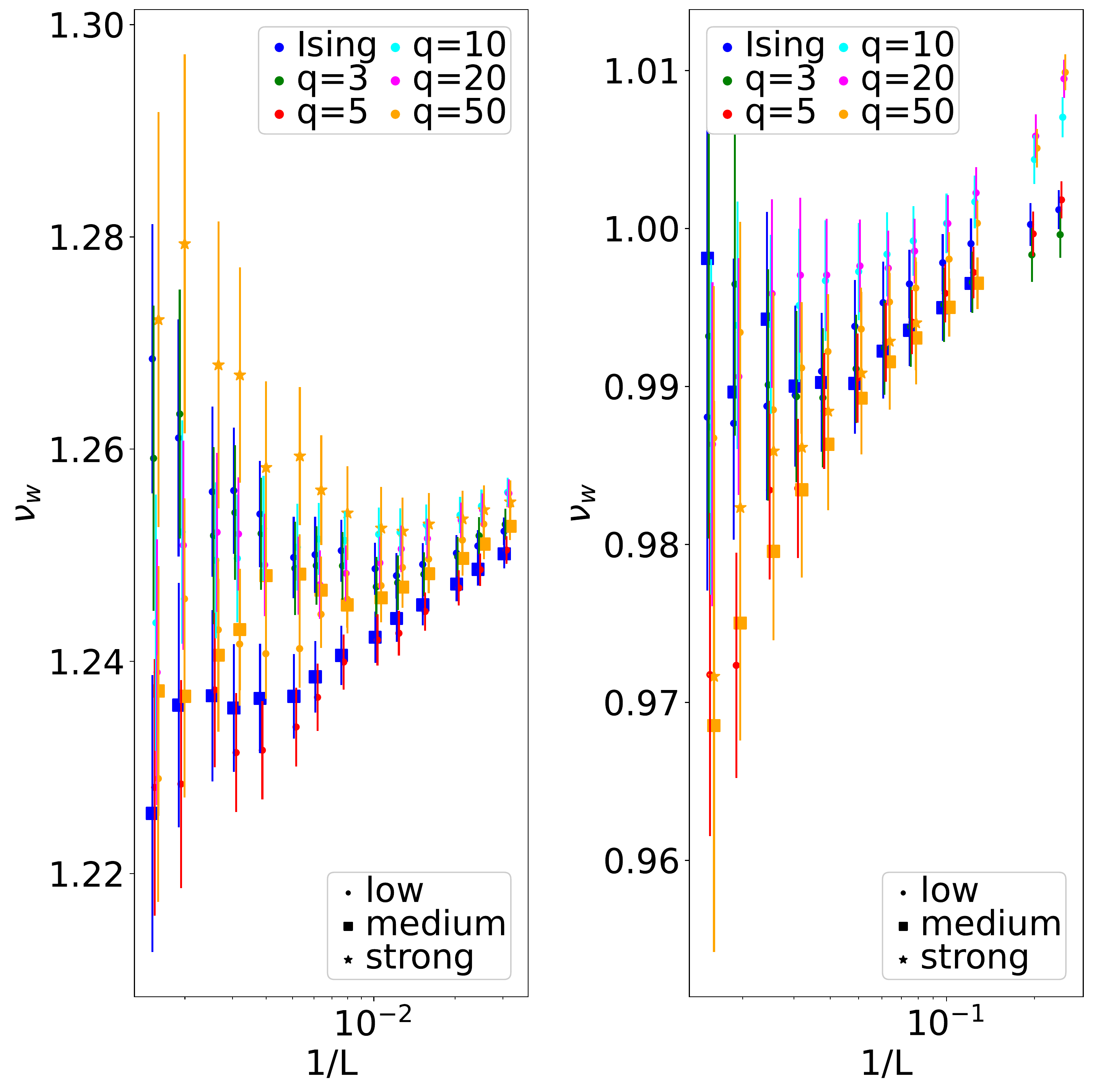}
    \caption{Effective exponent $\nu_w$ versus the smallest lattice size
    $L_{\rm min}$ entering into the fit for the 2D (left) and 3D (right)
    Potts models. The different colors correspond to different numbers of states
    $q$ of the Potts model. The different symbols correspond to different
    initial probability distributions of the couplings.}
    \label{fignuw2}
\end{figure}

Finally, a third estimate of the correlation exponent $\nu$ was obtained from
the standard deviation $\Delta\theta_c(L)=\sqrt{\overline{\theta_c^2(L)}
-(\overline{\theta_c(L)})^2}$. The latter is expected to scale as $L^{-1/\nu}$
with the lattice size. A linear fit (in log-log scale) with only two free parameters
is therefore sufficient in this case. To distinguish from the previous estimates,
the exponent estimated from the standard deviation $\Delta\theta_c(L)$ will be
denoted as $\nu_w$ in the following. An example for the 2D and 3D 10-state Potts
model is presented in Fig.~\ref{fignuw}.
The presence of scaling corrections is observed in the
3D case. Effective exponents $\nu_w(L_{\rm min})$, estimated by restricting the fit
to lattice sizes $L\ge L_{\rm min}$, are shown on Fig.~\ref{fignuw2}.
In the 2D case, the effective exponents $\nu_w$ are spread around the value $1.25$
for small $L_{\rm min}$, i.e. for fits over all or most of the lattice sizes,
and are compatible within error bars.  However, different evolutions are observed
as $L_{\rm min}$ is increased. Some exponents increase while others decrease.
Note that the effective exponents $\nu_w(L_{\rm min})$ for a given number of
states $q$ and a given initial distribution of the couplings are highly correlated
because they were computed with fits over the same set of data.
We were not able to identify a correlation between these different behaviors
and the number of states $q$ of the Potts model or the strength of disorder
in the initial distribution of the couplings. These different evolutions at
large $L_{\rm min}$ are therefore probably statistical fluctuations.
The 3D case is hopefully a bit clearer. The effective exponents decrease with
$L_{\rm min}$ and are compatible within error bars or close to the value
$\nu_w\simeq 0.98$ in the limit $1/L_{\rm min}\rightarrow 0$. We have no
reason to believe that different Potts models belong to different
universality class.
\\

From the effective exponents at large $L_{\rm min}$, a rough estimate of the
correlation lengths exponents can be inferred: $\nu_s\simeq 1.25(6)$ and
$\nu_w\simeq 1.25(3)$ in the 2D case, and $\nu_s\simeq 1.01(5)$ and
$\nu_w\simeq 0.985(10)$ in the 3D case. These estimates are compatible
 within error bars with the values of the literature for the Ising model:
$1.24(2)$ in the 2D case~\cite{Kovacs1} and $\nu_s=0.99(2)$ and $\nu_w=0.97(5)$
in the 3D case~\cite{Kovacs2}.

\section{Magnetic exponent}
The computation of the magnetic moment $\mu$ of the last decimated cluster
has revealed unexpected difficulties: for a given sample, $\mu$ often displays
a jump at the pseudo-critical point. This jump is rare and quite small for the
Ising model but more frequent and larger when the number of states $q$ of the
Potts model is increased. Since the pseudo-critical points $\theta_c^i(L)$ of
each sample was estimated with an accuracy of $10^{-5}$, the estimates of the
magnetic moment take randomly the value at the left or at the right of the
jump. The average magnetic moment is therefore equal to the mean of the values
at the left and at the right of the jump. We have checked that the relative
width of this jump decreases with the lattice size $L$.
For power-law initial distributions of the couplings (\ref{Proba2}),
broader than the uniform one, smaller jumps were observed.
\\

\begin{figure}
    \centering
    \includegraphics[width=0.47\textwidth]{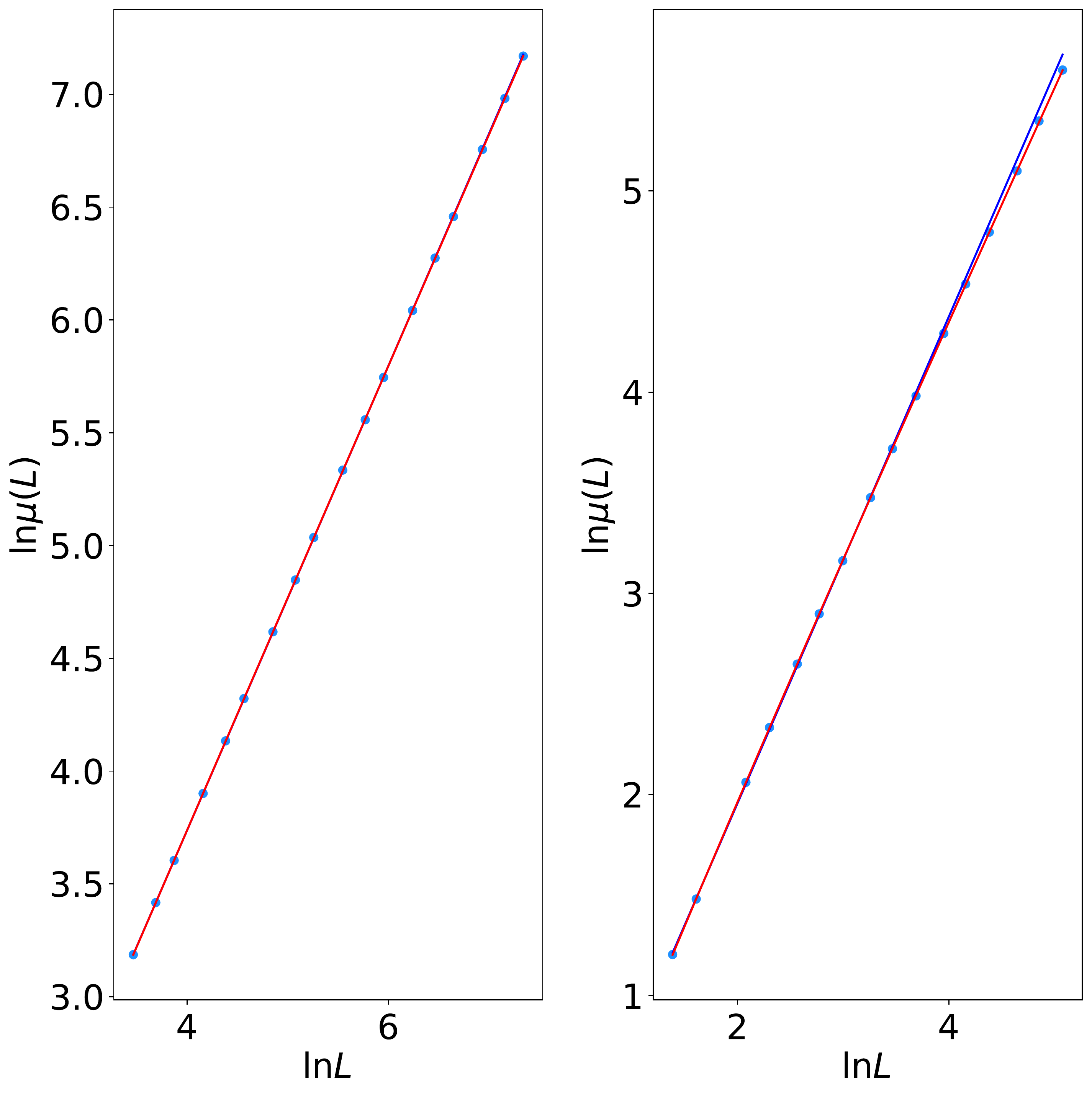}
    \caption{Average magnetic moment of the last decimated cluster $\bar\mu(L)$
    versus the lattice size $L$ of the 2D (left) and 3D (right) $q=10$ Potts model.
    In the 3D case, the light blue curve corresponds to a simple fit
    without any correction while the red curve corresponds to a fit with
    an algebraic correction.}
    \label{figdf}
\end{figure}

\begin{figure}
   \centering
    \includegraphics[width=0.47\textwidth]{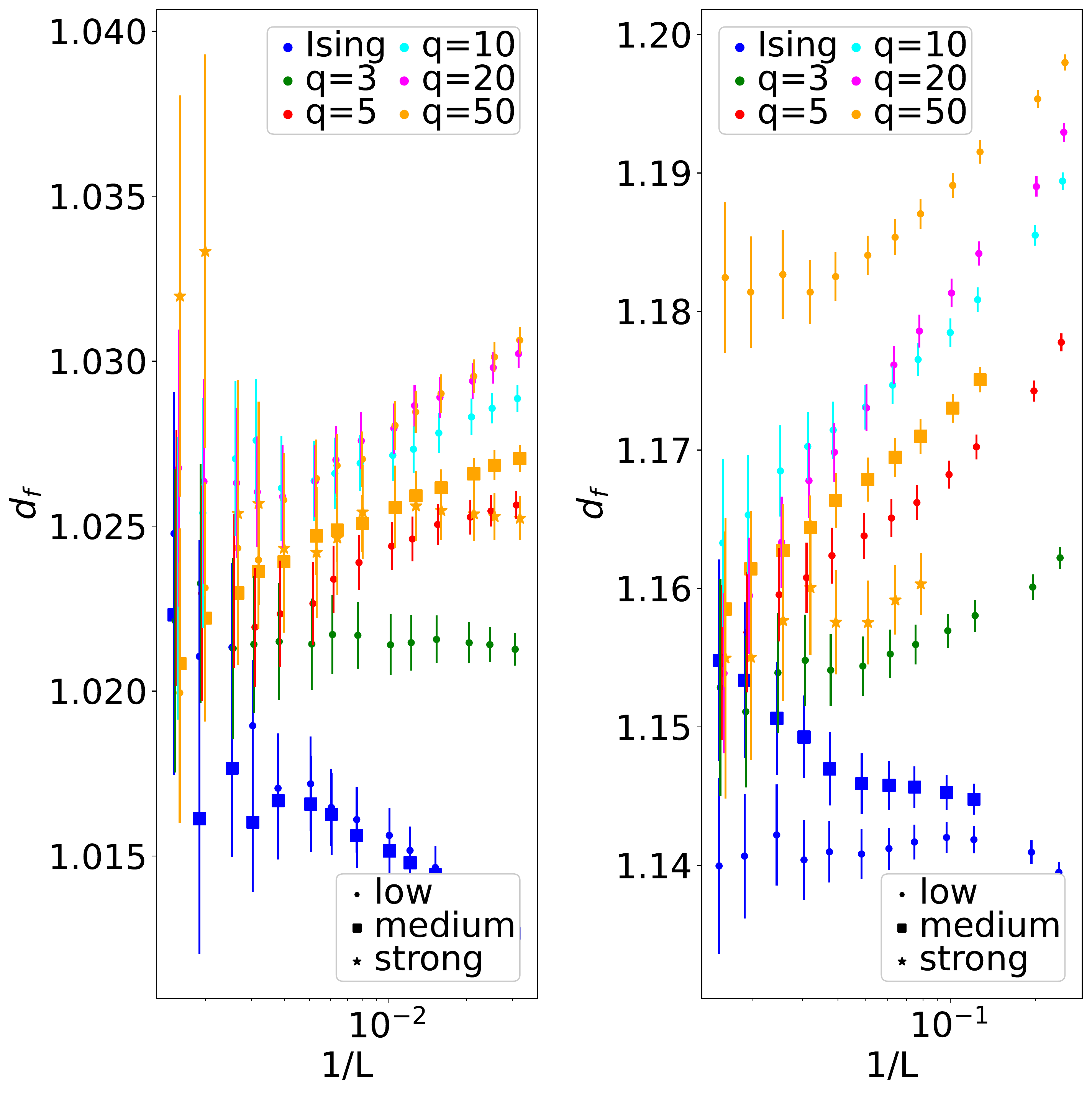}
    \caption{Effective exponent $d_f$ versus the smallest lattice size
    $L_{\rm min}$ entering into the fit for the 2D (left) and 3D (right)
    Potts models. The different colors correspond to different numbers of states
    $q$ of the Potts model. The different symbols correspond to different
    initial probability distributions of the couplings.}
    \label{figdf2}
\end{figure}

The fractal dimension $d_f$ of magnetization is estimated from the Finite-Size
Scaling
   \begin{equation}
    \bar\mu\sim L^{d_f}.
    \label{eq3}
    \end{equation}
of the average magnetic moment $\bar\mu$ of the last decimated
cluster at the pseudo-critical point $\theta_c^i(L)$ of each sample.
The average magnetic moment $\bar\mu$ is plotted in Fig.~\ref{figdf} for
the 2D and 3D 10-state Potts models. For the other values of the number of
states $q$ that were considered ($q=3, 5, 20$ and 50), the curves are similar.
The blue curve is the result of a linear fit of $\ln\bar\mu$ with $\ln L$.
In both the 2D and 3D cases, the estimated fractal dimensions $d_f$ are
incompatible and systematically increase with the number of states $q$ of the
Potts model. A non-linear fit with an algebraic correction, $\bar\mu=aL^{d_f}
\big(1+bL^{-\omega}\big)$ was performed but none of the algorithms that were used
gave a stable estimate of $d_f$. To nevertheless take into account possible
$q$-dependent scaling corrections, the fit was performed on various ranges
of lattice sizes. As for the correlation-length
exponent, a $L_{\rm min}$-dependent effective exponent $d_f(L_{\rm min})$ is
estimated by a fit restricted to the lattice sizes $L\ge L_{\rm min}$. This
effective exponent is shown in Fig.~\ref{figdf2}. The influence of a correction
is clearly seen as a dependence with $1/L_{\rm min}$. In the case of the 2D
random Potts model, the fractal dimension $d_f$ increases with $L_{\rm min}$
for the Ising model, is roughly stable for $q=3$ and decreases for $q\ge 4$.
As a consequence, the spreading of the effective exponents $d_f(L_{\rm min})$
decreases with $L_{\rm min}$. Even though all exponents are not compatible within
error bars, the super-universality of the 2D random Potts models seems much more
plausible than without taking into account scaling corrections.
In the case of the 3D random Potts model, the scenario is similar. When
scaling corrections are not taken into account, the estimates of the fractal
dimensions $d_f$ are fully incompatible. As $L_{\rm min}$ is increased, the
spreading of the effective exponents $d_f(L_{\rm min})$ shrinks and the
fractal dimensions are  finally compatible with the
value $1.155$. Two exceptions can be noticed: the Ising model and the $q=50$
Potts model with weak disorder, whose exponents are still far from $1.155$
at large $L_{\rm min}$. However, at medium disorder, the exponents
are remarkably closer to others and are compatible with $1.155$
at large $L_{\rm min}$.

From the effective exponents at large $L_{\rm min}$, the fractal dimensions
can be estimated to be $d_f\simeq 1.021(5)$ for the 2D random Potts model
and $1.155(8)$ for the 3D model. These estimates are compatible within error
bars with the values of the literature for the Ising model: $1.018(16)$ for
the 2D Ising model~\cite{Kovacs1} and $1.161(15)$ for the 3D Ising
model~\cite{Kovacs2}.

\section{Energy excitation}
\begin{figure}
    \centering
    \includegraphics[width=0.47\textwidth]{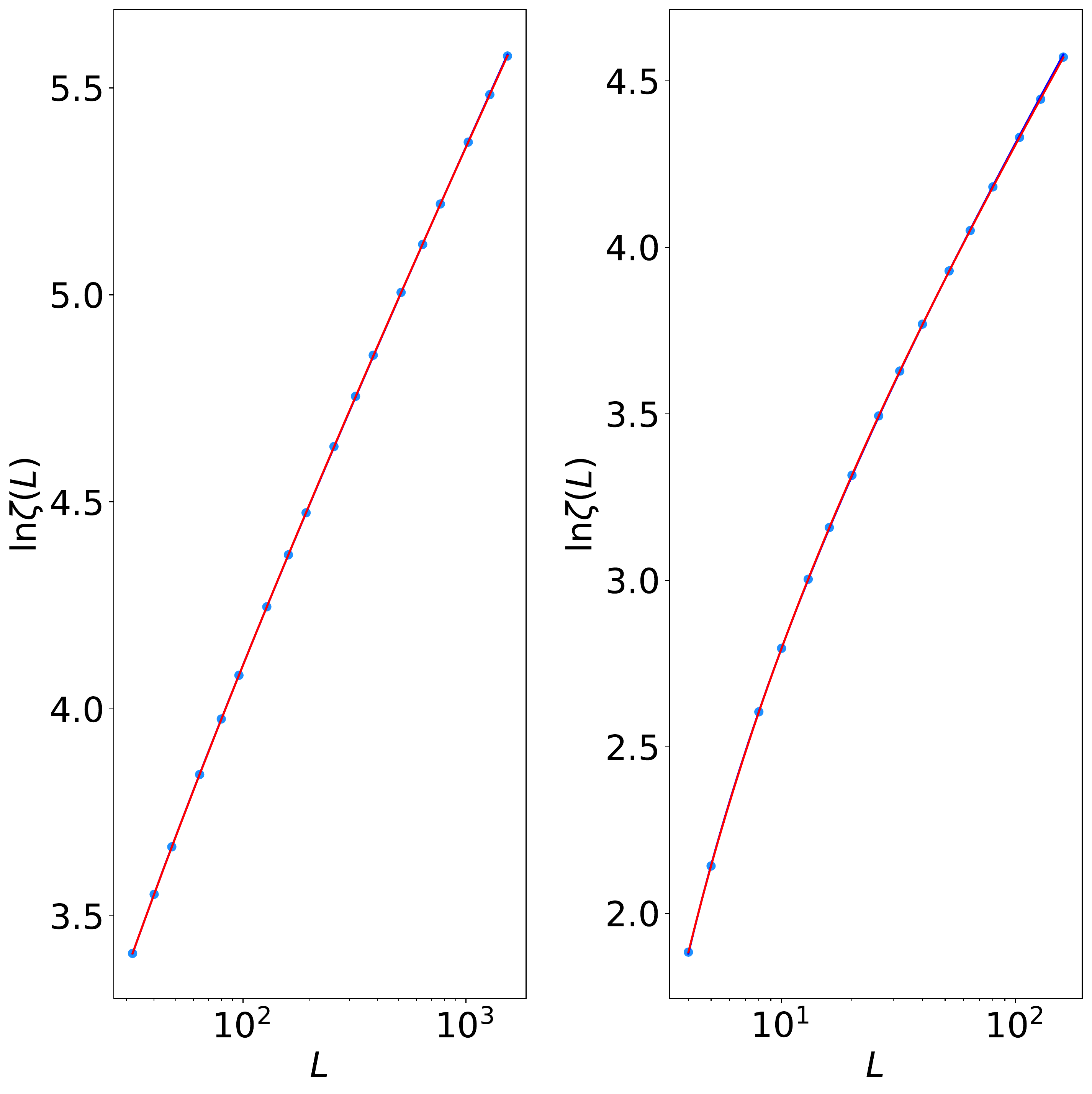}
    \caption{Average energy gap $\zeta=-\ln\overline{\Delta E}$ of the last
    decimated cluster versus the lattice size $L$ of the 2D (left) and 3D
    (right) $q=10$ Potts model. The curve is the non-linear fit
    $-\ln\overline{\Delta E}=a+bL^{\psi}$.}
    \label{figpsi}
\end{figure}

\begin{figure}
   \centering
    \includegraphics[width=0.47\textwidth]{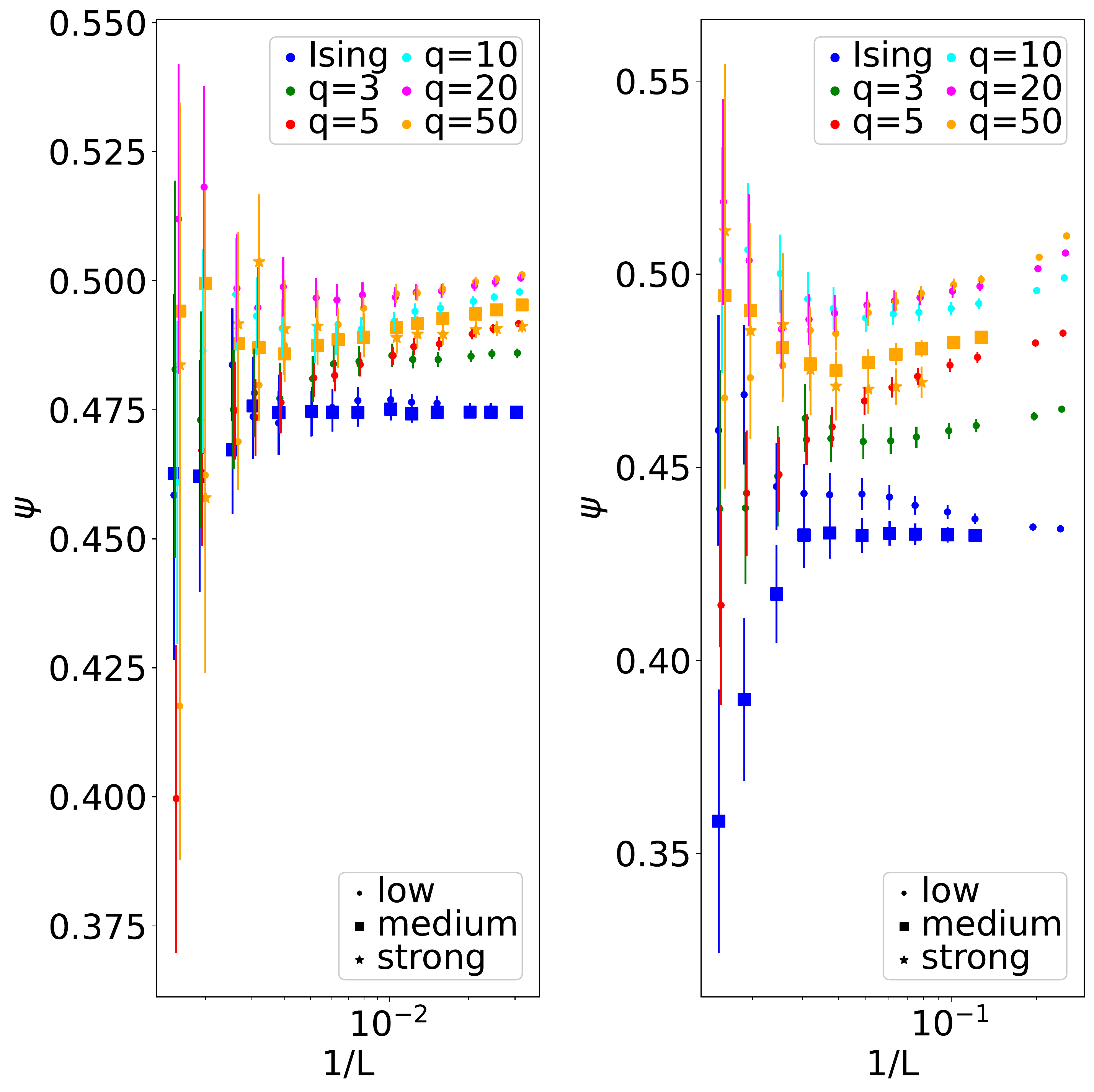}
    \caption{Effective exponent $\psi$ versus the smallest lattice size
    $L_{\rm min}$ entering into the fit for the 2D (left) and 3D (right)
    Potts models. The different colors correspond to different numbers of states
    $q$ of the Potts model. The different symbols correspond to different
    initial probability distributions of the couplings.}
    \label{figpsi2}
\end{figure}

At the IDFP, the largest energy scale $\Omega$ decreases as
    \begin{equation}
    \Omega\sim \Omega_0e^{-cL^{\psi}}\ \Leftrightarrow\
    \log{\Omega\over\Omega_0}\sim -L^{\psi}
    \label{eqpsi}
    \end{equation}
during the SDRG flow. The same relation is expected to hold for the energy gap
$\Delta E$ of the last decimated cluster, i.e. the smallest transverse field
that was decimated in our implementation of the SDRG. On Fig.~\ref{figpsi}, the
numerical data are presented for the 10-state 2D and 3D random Potts models. A non-linear
fit of $-\ln\overline{\Delta E}=a+bL^{\psi}$ is performed over the lattice sizes
$L\ge L_{\rm min}$ to estimate an effective critical exponent $\psi(L_{\rm min})$.
The latter is shown in Fig.~\ref{figpsi2}. Note that there are 3 free parameters
in this fit so the accuracy will be smaller than for $\nu_w$ and $d_f$.
In the 2D case, the effective
exponents $\psi(L_{\rm min})$ are incompatible and increase with the number
of states $q$ of the Potts model for small $L_{\rm min}$. However, for large
$L_{\rm min}$, the $q$-dependence becomes smaller and the exponents
are finally compatible within error bars with the value $0.485$.
In the 3D case, the exponents $\psi$ are also incompatible and increase with
the number of states $q$ of the Potts model at small $L_{\rm min}$. As in the
2D case, the spreading shrinks at large $L_{\rm min}$ but not enough for the
exponents to become compatible with error bars. Larger lattices sizes would
be helpful to reach a definitive conclusion.

From the effective exponents at large $L_{\rm min}$, the exponent $\psi$
can be estimated to $\psi\simeq 0.48(2)$ for the 2D random Potts model and
$0.46(4)$ for the 3D model. These estimates are compatible within error
bars with the values of the literature for the Ising model:
$0.46(2)$ for the 2D Ising model~\cite{Kovacs1} and $0.48(2)$ for the 3D
Ising model~\cite{Kovacs2}.

\begin{table}
\caption{\label{tab2}Summary of the estimates of the critical
exponents of the 2D and 3D random Potts models.}
\begin{ruledtabular}
\begin{tabular}{l|llll}
 & $\nu_s$ & $\nu_w$ & $d_f$ & $\psi$ \\
\hline
2D & $1.25(6)$ & $1.25(3)$ & $1.021(5)$ & $0.48(2)$ \\
3D & $1.01(5)$ & $0.985(10)$ & $1.155(8)$ & $0.46(4)$ \\
\end{tabular}
\end{ruledtabular}
\end{table}

\section{Conclusions}
When scaling corrections are not taken into account, the critical exponents
$d_f$ and $\psi$ were shown to take, both in 2D and 3D, incompatible values
that increase with the number of states $q$ of the Potts model.
Since non-linear fits involving algebraic scaling corrections are
unstable, effective exponents were estimated over shrinking ranges
of lattice sizes $L\ge L_{\rm min}$. The limit $1/L_{\rm min}\rightarrow 0$
of these effective exponents is expected to give the critical exponent at the
IDFP. The analysis is however hampered by the large fluctuations
of the effective exponents, due to the smaller number of degrees of
freedom in the fit at large $L_{\rm min}$. Nevertheless, scaling
corrections are clearly observed. While the latter do not seem to
depend on $q$ for the correlation length exponent (apart from $\nu_s$ in
the 3D case), the corrections
on the magnetic fractal dimension $d_f$ are strongly $q$-dependent.
A shrinking of the dispersion of the effective exponents for
different numbers of states $q$ is observed when $L_{\rm min}$ is
increased, providing evidence of the existence of a super-universality
class. This conclusion is in agreement with Ref.~\cite{Vasseur} where
the super-universality of the 2D random Potts model is shown by means
of a mapping onto a lattice gauge model.
The effective exponents $\psi$ display a behavior similar to $d_f$,
although less pronounced and with a residual dispersion of the
exponents in the limit $1/L_{\rm min}\rightarrow 0$.
Our final estimates of the critical exponents
at the IDFP are summarized in Tab.~\ref{tab2}. As noticed in the text,
they are compatible with previous estimates obtained for the Ising model.
\\

The question of the extension of the random quantum Ising universality
class naturally arises. The 1D case was discussed in the introduction.
In dimensions $d\ge 2$, it is not known whether the Ashkin-Teller model
or the clock model also belong to the Ising universality class? There
are some of the open questions that we will try to address in the future.

\begin{acknowledgments}
The numerical simulations of this work were performed at the meso-center
eXplor of the universit\'e de Lorraine under the project 2018M4XXX0118.
\end{acknowledgments}

\end{document}